\documentclass[conference]{IEEEtran}
\IEEEoverridecommandlockouts
\usepackage{cite}
\usepackage{amsmath,amssymb,amsfonts}
\usepackage{algorithmic}
\usepackage{graphicx}
\usepackage{textcomp}
\usepackage{xcolor}
\def\BibTeX{{\rm B\kern-.05em{\sc i\kern-.025em b}\kern-.08em
    T\kern-.1667em\lower.7ex\hbox{E}\kern-.125emX}}
\begin{document}

\title{Recognizing Human Internal States: A Conceptor-Based Approach\\}

\author{\IEEEauthorblockN{1\textsuperscript{st} M E Bartlett}
\IEEEauthorblockA{\textit{University of Plymouth}\\
Plymouth, United Kingdom}
\and
\IEEEauthorblockN{2\textsuperscript{nd} Daniel Hern\'andez Garc\'ia}
\IEEEauthorblockA{\textit{University of Plymouth}\\
Plymouth, United Kingdom}
\and
\IEEEauthorblockN{3\textsuperscript{rd} Serge Thill}
\IEEEauthorblockA{\textit{University of Sk\"ovde}\\
Sk\"ovde, Sweden\\
\textit{Donders Institute}\\
Nijmegen, The Netherlands}
\and
\IEEEauthorblockN{4\textsuperscript{th} Tony Belpaeme}
\IEEEauthorblockA{\textit{University of Ghent}\\
Ghent, Belgium\\
\textit{University of Plymouth}\\
Plymouth, United Kingdom}
}

\maketitle

\begin{abstract}
The past few decades has seen increased interest in the application of social robots to interventions for Autism Spectrum Disorder as behavioural coaches \cite{Scass12}. We consider that robots embedded in therapies could also provide quantitative diagnostic information by observing patient behaviours. The social nature of ASD symptoms means that, to achieve this, robots need to be able to recognize the internal states their human interaction partners are experiencing, e.g. states of confusion, engagement etc. Approaching this problem can be broken down into two questions: (1) what information, accessible to robots, can be used to recognize internal states, and (2) how can a system classify internal states such that it allows for sufficiently detailed diagnostic information? In this paper we discuss these two questions in depth and propose a novel, conceptor-based classifier. We report the initial results of this system in a proof-of-concept study and outline plans for future work.
\end{abstract}

\begin{IEEEkeywords}
Internal States, Engagement, Conceptors, Socially Interactive Robots, Recognition
\end{IEEEkeywords}

\section{Introduction}

The development of socially interactive robots has inspired research into various applications for these tools. One application is in therapy and care, where robots can be used to provide daily support to patients, and as tools to augment interventions and provide quantitative data for clinicians \cite{van16}. We specifically consider the use of robots in interventions for children with Autism Spectrum Disorder (ASD). The Diagnostic and Statistical Manual of Mental Disorders (DSM-V) defines ASD as a neuro-developmental disorder characterized by persistent deficits in social communication and interaction, and restricted or repetitive behaviours and interests \cite{APA13}. Diagnosing ASD involves the subjective interpretations by experts of observations of a child's behaviour made by clinicians and caregivers \cite{Rog16}. This subjectivity, and the clinical heterogeneity typical between ASD cases \cite{Scass12}, means that the diagnostic process could be improved through the use of more quantitative, objective measures of child behaviour. This can be achieved using behaviour classification systems. 

Developing an artificial system to recognize ASD symptoms is not a straight-forward task due to the social nature of ASD. This is because correct classification of social and interaction behaviour often requires the ability to infer the internal-states (e.g. intentions, emotions) of the observed individual. For example, identifying when a child fails to ask for comfort when needed (a symptom of ASD \cite{APA13}) requires that the observer recognize that the child is experiencing a negative internal state. However, endowing robots with this skill would provide numerous benefits for ASD interventions. For instance, if an intervention involves regular interaction with a social robot, it would be useful to have the robot able to report quantitative diagnostic information. Firstly, clinicians could use this information to track their patient's progress through the intervention, or to support their initial diagnostic decision. Secondly, the robot itself could use internal-state and diagnostic information to autonomously decide on appropriate behaviours to perform.  

In approaching the problem of developing artificial systems able to recognize human internal states, there are two key questions which must be addressed: (1) what internal state information is available in behaviours which can be assessed and quantified by artificial systems, and (2) how can these states be represented by a classification system to provide both detailed assessments and flexible behavioural responses from a social robot. The rest of this paper discusses possible answers to these questions in the context of quantifying the diagnostic behaviours of children with ASD. We present two studies carried out as a proof-of-concept to demonstrate that the internal state of task engagement could be classified based on observable human movement information, and that this classification could be done by a system able to represent internal states as points along a continuous dimension. The logic behind our choice of internal state and its desired representation is described, where relevant, in the introductions to each experiment. 

\section{Experiment 1}

Whilst most ASD symptoms cannot be described as wholly overt, many have been linked with directly observable behaviours. For example, motor skills have been shown to be predictive of social communication skills for children with ASD \cite{Brad18}. Additionally, an increased tendency to orient towards non-social contingencies rather than biological motion is indicative of ASD \cite{klin09}. These and other studies have linked movement and gaze behaviours to several ASD characteristics. Movement and gaze information can be measured or estimated by observing body movements or poses, which can be easily made accessible to artificial systems, e.g. by converting the position of an individual's joints to coordinates in space. Consequently, we argue that such observable information can be useful for social robots designed to make inferences about human internal states pertaining to ASD symptoms. 

Designing a system to recognize this kind of diagnostic information, however, is non-trivial. We would need to have identified how observable behaviours relate to symptomology, and define which symptoms we are best able to recognize and describe in terms of severity based on behavioural data. Given the complexity of obtaining and labelling such data, we chose to perform a proof-of-concept study demonstrating the feasibility of our approach using data from a non-clinical population. We therefore chose to examine whether the internal state of task engagement could be identified and classified into different classes, based on the `intensity' of the experienced state. That is, we aimed to train a classifier to distinguish between `high', `intermediate' and `low' task engagement based on the behaviour of typically developing children. Before a classifier could be implemented, however, we first needed to verify that the internal state of interest (task engagement) was recognizable from the movement information available in our data set.  

For this study, the desired data set was defined as one which contained the movement information of humans experiencing, but not intentionally communicating, different levels of a non-emotional internal state. To ensure that the internal state was not being communicated we decided that the subject should not be interacting with another human. With these considerations in mind, the data set for this experiment was taken from the openly available PInSoRo data set\cite{lem17}\footnote{https://freeplay-sandbox.github.io}. This data set comprises videos of child-robot pairs interacting with each other and a touch-screen table-top (the sand-tray). We argue that these videos meet the requirements of showing humans experiencing internal states which could be described along a continuum (i.e. engagement with the touch-screen) which were not being actively communicated (i.e. due to the lack of a human interaction partner). The videos have been annotated for a number of behaviors including whether the child was engaged in ``goal oriented'', ``aimless'' or ``no'' play. We believe these annotations are analogous to ``high'', ``intermediate'' and ``low'' levels of task engagement respectively. A preliminary study was designed to validate this assumption. 

\subsection{Method}
\subsubsection{Participants}
Five participants (students and employees) were recruited from the University of Plymouth's School of Computing, Electronics and Mathematics on a volunteer basis. Demographic information was not collected. 

\subsubsection{Materials}
A total of forty-five video clips were extracted from the data set for this study. We selected fifteen clips with the annotation ``goal-oriented play'', fifteen with the annotation ``aimless play'' and fifteen with the annotation ``no play''. Clip lengths ranged from 12-30 seconds. 

After clips were selected we extracted both the full visual scene versions and the movement-alone versions. The movement-alone versions were processed such that they depicted the children's joint-points, connected by coloured lines, against a black background. These videos act as visual representations of the data used as input for the conceptor-based system in that they depict only movement and pose information by showing the position of the child's body in each frame. 

\subsubsection{Procedure}
For each participant the experiment was conducted over two days. Participants watched the full visual scene videos on the first day and were then asked to return the next day when they would watch the movement-alone videos. Participants all received the following instructions before beginning the experiment:
\begin{center}
    \textit{You're about to watch several videos of children interacting with a touch-screen table-top. The children were able to either play a specific game on the touch-screen, or to do whatever they want. After each clip you will be asked to judge the child's level of task engagement.}
\end{center}
Participants were then given the opportunity to ask any questions they may have had and were instructed about their right to withdraw before beginning the experiment.

This study was created using JSPsych and presented on a desktop computer. Participants were positioned a comfortable distance away from the screen where they could still reach the keyboard and mouse to provide responses. At the beginning of the experiment, the instructions were reiterated. Participants were then presented with a consent form within the experiment script and given two response options. If participants selected the ``I consent'' option, the experiment proceeded as normal. If participants selected ``I do not consent'' the experiment was terminated. Participants then viewed nine of each type of clip (a total of twenty-seven clips) presented in a random order. Following each clip, participants were presented with the question ``\textit{How engaged was the child with their task on the touch screen table-top?}''. This question was accompanied by a 7-point Likert scale ranging from 1 = ``Not at all Engaged'' to 7 = ``Highly Engaged''. Participants used this scale to report how engaged they thought the child in the clip had been and then continued on to the next clip. 

At the end of the experiment on the first day, participants were given the opportunity to ask any questions they had and were asked to return the next day to complete the second half. On the second day, the experiment proceeded in the same way except participants were shown the movement-alone videos instead of the full visual scene videos. Each participant saw the same twenty-seven clips in both sessions. At the end of the second session participants were fully debriefed on the nature and purpose of the study and were thanked for their participation. Each session took approximately 10-15 minutes to complete. 

\subsection{Results}
The following analyses were run using RStudio. 

\subsubsection{Inter-Rater Agreement}
The data were analyzed in two main ways. We firstly examined inter-rater agreement by calculating Krippendorff's alpha for the responses. We initially checked whether participants gave similar responses for each of the three types of videos. To do this, Krippendorff's alpha was calculated for responses to all of the videos of each type. The alpha scores have been interpreted in terms of the benchmarks outlined by Landis and Koch \cite{land77}. Responses showed ``fair'' agreement for the goal-oriented (high engagement) clips (Krippendorff's alpha = 0.269) and the no-play (low engagement) clips (Krippendorff's alpha = 0.267). Responses for aimless (intermediate engagement) clips showed ``slight'' agreement (Krippendorff's alpha = 0.171). The low levels of agreement can partially be explained by the fact that there were very few raters (2-4) per clip. As such we did not expect perfect levels of agreement and argue that the levels obtained suggest a sufficient degree of similarity in participants' ratings. 

We then examined whether participants had higher agreement when viewing the full visual scene clips compared to the movement-alone clips for each clip type. The results of this analysis are reported in Table 1. For the goal-oriented and no-play clips, participants tended to show similar levels of agreement in each condition. However, for the aimless clips, participants demonstrated poor agreement when viewing the movement-alone clips. 


\begin{table}[h!]
    \centering
    \caption{Table of inter-rater agreement scores for responses to each clip-type in each condition}
    \begin{tabular}{|p{3cm}|p{2cm}|p{2cm}|}
         \hline
         Clip Type & \multicolumn{2}{c|}{Krippendorff's Alpha (3 d.p.)}  \\ [0.5ex]  
         \hline
         & Full Scene & movement-alone \\ 
         \hline
         Goal Oriented & 0.382 (fair) & 0.368 (fair)\\
         \hline
         Aimless & 0.247 (fair) & -0.022 (poor) \\ 
         \hline
         No Play & 0.126 (slight) & 0.202 (fair) \\ 
         \hline
    \end{tabular}
\end{table}

\subsubsection{Ratings} 
The second set of analyses looked at the how participants rated each type of video. Overall mean rating was 4.81 (SD $=$ 1.25) for goal-oriented clips, 4.16 (SD $=$ 1.52) for aimless clips, and 2.43 (SD $=$ 1.54) for no-play clips. An ANOVA revealed a significant main effect of clip-type on ratings (F(2,267)=64.99, p$<$0.001). Importantly, a \textit{post hoc} Tukey test revealed significant differences between all conditions (Tukey’s HSD: all differences $>$0.6, all p’s $<$0.007; see Table 2). 

\begin{table}[h!]
    \caption{Table of results for post hoc Tukey's Honest Significant Difference test.}
    \begin{tabular}{|p{3cm}|p{1.5cm}|p{2.5cm}|}
         \hline
         Comparison & Difference & Significance (p adj)  \\ [0.5ex]  
         \hline
         Goal Oriented $-$ Aimless & 0.656 & $p=0.007$ \\
         \hline
         Goal Oriented $-$ No Play & 2.348 & $p<0.001$\\ 
         \hline
         Aimless $-$ No Play & 1.722 & $p<0.001$\\ 
         \hline
    \end{tabular}
\end{table}

These results demonstrate that participants rated the clips in terms of engagement such that goal-oriented clips showed the highest levels of engagement, no-play clips showed the lowest levels, and aimless clips fell in-between these two extremes. Consequently, we feel our assumption that these annotations reflect different levels of engagement is sufficiently supported for these data to be used to train and test a conceptor-based classifier to recognize engagement based on observable behaviour. The remainder of this paper describes the design and initial tests of such a classifier. 

\section{Experiment 2}
In addressing the second question of how to represent internal states, we consider that ASD diagnosis involves ranking behaviours in terms of severity \cite{lord2000}. In this way, behaviours important to ASD diagnosis can be thought of as lying along a continuum of severity. To emulate this we need a classification technique which can identify different `levels' along a continuous dimension. This can be achieved using classical machine learning techniques by training a classifier on examples of each severity level. However, obtaining a large enough training data set for this would be very time-consuming and difficult, owing to the need to have expert commentators provide a severity label for each example. We therefore require a method which can learn several classification categories for each behaviour of interest, using a limited training data set. One approach which is suited to this task is conceptors \cite{Jae14}. 

Conceptors are neuro-computational mechanisms that can be used for learning a large number of dynamical patterns based on learned prototypical extremes \cite{Jae14}. This approach assumes that there is a continuum underlying the behavior. New patterns can be generated by combining and morphing the learned extremes. As such, we argue that conceptors may be appropriate for classifying human internal states. The second study described here tested this hypothesis by designing a conceptor-based system to recognize task engagement from observable human movements.

\subsection{Method}
\subsubsection{Materials}
The data set for this study was again taken from the PInSoRo data set. All of the clips annotated with the labels ``goal-oriented play'' (high engagement) and ``no play'' (low engagement) were extracted (total of 354 clips). Clips were preprocessed such that the xyz coordinates of the child's joints in each frame were taken as the input for the conceptor-based classifier. A subset of ``high'' (62 clips) and ``low'' (115 clips) engagement clips made up the training data set. The remaining 177 clips made up the test data set. 

\subsubsection{Conceptor-Based Classifier}

The conceptor-based approach is based on a key dynamical phenomenon in Recurrent Neural Networks; ``if a `reservoir' is driven by a pattern, the entrained network states are confined to a linear subspace of network state space which is characteristic of the pattern'' \cite{Jae14}. In this way the dynamics of a pattern (in our case an overt behavior for a classifiable activity like engagement) will occupy different regions of the state space, and can be encoded in a conceptor. A conceptor ($ C_{j}$) acts as a map associated with a pattern ($p_{j}$). To build a conceptor-based classifier we computed $J$ conceptors, one for each class in our classifier. To obtain the conceptors an echo state network (ESN) was first created with an input layer of $K$ input units and a hidden layer reservoir of $N$ neurons. For each class the network will be driven, independently, with all training samples $s_{j}^{m}$ in each class $j$, according to the ESN state update equation:
\begin{equation}
  x(n+1) = \tanh(W \cdot x(n) + W^{in} \cdot p(n+1) + b)
  \label{eq:update}
\end{equation}
This yielded a set of network states $X_{j}= [x(1) \dots x(t)]$ where $t$ is the number of time-steps in $s_{j}$ from which a state correlation matrix $ R_{j} = X_{j}X_{j}^{T} / M_{j} $ is obtained, where $ M_{j} $ is the total number of samples for class $j$. Next we computed conceptor $C_{j}$ through the equation:
\begin{equation}
  C(R, \alpha) = R(R+\alpha^{-2})^{-1}
  \label{eq:conceptor}
\end{equation}
Where $R$ is a correlation matrix and $\alpha \in (0, \infty)$ in an ``apperture'' parameter. For more see \cite{Jae14}.

Once we computed a conceptor matrix for each class we were able to classify a new sample $s$ from the test set by feeding it into the ESN reservoir to obtain a new state vector $z = [x(1) \dots x(n)] $. then, for each conceptor, the ``positive evidence'' quantity $z^{T}C_{j}z$ was computed. This led to a classification by deciding for $j = argmax(z^{T}C_{j}z)$ as the class $j$ to which the sample $s$ belongs.
The procedure for the conceptor-based classifier is summarize in Table \ref{tab:classifier}.
\begin{table}
\begin{center}
{\footnotesize
\caption{Predicting internal states with Conceptors.}
\begin{tabular}{ p{200pt} }
    \textbf{Algorithm:} Conceptor-based classification.  \tabularnewline  
    \textbf{Input:} A test sample $s$ belonging to one a class $j$. \tabularnewline
	\begin{enumerate}
		\item Take a sample $s$ from the test set.
		\item Drive the reservoir with sample $s$ to obtain a state vector $z = [x(1) \cdots x(n)] $, where $n$  is the $\#$ of steps in $s$.
		\item For each Conceptor $C_{j}$ compute $h(j) = z^{T}C_{j}z$, a ``positive evidence'' quantity of $z$ belonging to class $j$.
		\item Collect each evidence $h(j)$ into a $j-$dimensional classification hypothesis vector $h^{+} = \{ h(1) \cdots h(j) \}$.
		\item Classify $s$ as belonging to class $j$ from $j = argmax(h^{+})$.
		\item \textbf{END}
	\end{enumerate}  \tabularnewline
	\textbf{Output:} Class sample $s$ belongs to. \tabularnewline
\end{tabular}
}
\label{tab:classifier}
\end{center}
\end{table}

\subsection{Results}
The resultant conceptors were tested using previously unseen samples from the high and low engagement categories. The results of this test are shown in Figure 1. Performance is above chance for both classes (high engagement: 60\%, low engagement: 75\%). 

\begin{figure}[htbp]
\centerline{\includegraphics[scale=0.5]{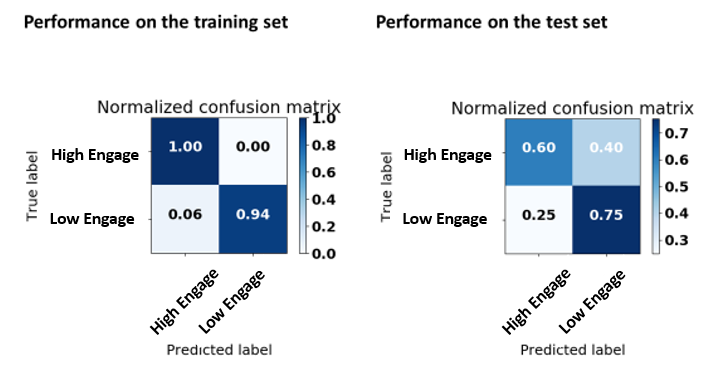}}
\caption{Confusion matrices showing classification performance of trained conceptors on training data (left) and test data (right).}
\label{fig}
\end{figure}

\section{Conclusions}
This study demonstrates that it is possible to train a conceptor-based system, on real non-periodic data, to classify between high and low engagement based on observable human behavior. The conceptor-based system successfully learned to recognize high and low engagement from observable human movement. Future work will construct new conceptors by linearly combining these learned conceptors. We will then test whether these new conceptors can be used to recognize intermediate levels of engagement identified in the PInSoRo data set.  

If new conceptors can be generated, this method will show promise for use in providing diagnostic information for clinicians assessing children with ASD. The ability to interpolate between extremes along a continuum means that such a system could be trained on a smaller dataset, whilst still achieving a high level of detail through the generation of multiple intermediate classification categories. 

\section*{Acknowledgment}
This work is part of the EU FP7 project DREAM project (www.dream2020.eu, grant nr. 611391) and the H2020 L2TOR project (www.l2tor.eu, grant nr. 688014).

\end{document}